\documentclass[pra,twocolumn,longbibliography,nofootinbib,floatfix]{revtex4-1}

\usepackage{balance}
\usepackage[utf8]{inputenc}
\usepackage{bm,amsfonts,amsmath,amssymb}
\usepackage{dcolumn,xfrac}
\usepackage{natbib}
\usepackage{easyReview}
\usepackage{tipa}
\usepackage{booktabs}
\usepackage{multirow}
\newcommand{\Eref}[1]{Eq.~(\ref{#1})}
\newcommand{\Tref}[1]{Table~\ref{#1}}

\usepackage{mathtools}
\makeatletter
\newcommand{\vast}{\bBigg@{4}}
\newcommand{\Vast}{\bBigg@{4.5}}
\makeatother

\newcommand{\dashrule}[1][black]{%
  \color{#1}\rule[\dimexpr.5ex-.2pt]{4pt}{.4pt}\xleaders\hbox{\rule{4pt}{0pt}\rule[\dimexpr.5ex-.2pt]{4pt}{.4pt}}\hfill\kern0pt%
}

\begin{document}
\title{Atomic calculations of hyperfine structure anomaly in gold}
\author{Yu. A. Demidov$^{1,2}$}
\author{E. A. Konovalova$^{1}$}
\author{R. T. Imanbaeva$^{1,2}$}
\author{M. G. Kozlov$^{1,2}$}
\author{A. E. Barzakh$^{1}$}
\affiliation{$^1$ Petersburg Nuclear Physics Institute NRC ``Kurchatov Institute'', 188300 Gatchina, Russia \\
$^2$ St.~Petersburg Electrotechnical University ``LETI'', 197376 St. Petersburg, Russia}

\begin{abstract}
The magnetic hyperfine structure constants have been calculated for low-lying levels in neutral gold atom and gold-like ion of mercury taking into account Bohr--Weisskopf (BW) effect. BW effect is represented as a product of atomic and nuclear ($d_\mathrm{nuc}$) factors. We have calculated the atomic factors, which enable one to extract BW-correction values for far from stability gold nuclei from the experimental data. The possible uncertainty of our atomic calculations have been estimated by the comparison with the available experimental data. It has been shown that the standard single-particle approach in $d_\mathrm{nuc}$ calculation reasonably well describes experimental data for $11/2^-$ gold isomers and $3/2^+$ ground state of $\rm ^{199}Au$. At the same time, it fails to describe the hyperfine constant in $^{197}\mathrm{Au}$. This indicates the more pronounced configuration mixing in $\rm ^{197}Au$ than in $\rm ^{199}Au$.
\end{abstract}
\maketitle

\section{Introduction}
Laser spectroscopy has become a very efficient tool to study the nuclear $g$-factors (the ratio of nuclear magnetic moment $\mu$ and nuclear spin $I$) of short lived isotopes and isomers. The $g$ factor of the isotope in question (marked by a star below) is usually extracted from the experimentally measured magnetic hyperfine structure (HFS) constant $A$ by the relation:
 \begin{align} \label{g_unst1} 
g^*_I = g^\mathrm{ ref}_I \frac{A^*}{A^\mathrm{ref}},
 \end{align}where superscript ``ref'' denotes a reference isotope with known $g$-factor and $A$-constant values. \Eref{g_unst1} is valid for a point-like nucleus only. The finite size of the nucleus leads to the deviation of the $A$ constant from the point-like value due to the distribution of magnetisation and charge over the nuclear volume. These corrections break proportionality between the hyperfine constants and nuclear $g$ factors and can be taken into account by introducing so called relative hyperfine anomaly (RHFA)~\cite{BW50}. For isotopes (1) and (2) the RHFA $^1\Delta^2$ is defined as:
 \begin{align} \label{HFAdef}
 ^1\Delta^2 \equiv   \frac{g_I^{(2)}A^{(1)}}{g_I^{(1)}A^{(2)}}-1.
 \end{align}
 The RHFA value can be found from \Eref{HFAdef} if nuclear $g$ factors were determined by experimental methods which don't rely upon HFS measurements. Therefore information about RHFA is limited primarily to the stable and long-lived isotopes where such methods can be applied. Experimentally measured RHFA values are as a rule within the range of $10^{-2}\, -\, 10^{-4}$~\cite{Per13}. Correspondingly, this correction is typically neglected and the uncertainty of the extracted $g$ factor is increased by $\approx 1$\%. In most cases this approach is acceptable in view of the experimental uncertainties and nuclear physics inferences. However, the large RHFA values \mbox{$\rm ^{197}\Delta^{198}= 8.53(8)\%$} and  \mbox{$\rm ^{197}\Delta^{180}= 21(14)\%$} were reported in Refs.~\cite{Ekstr80, harding20}. It exceeds the experimental uncertainties of HFS constants in gold isotopes by orders of magnitude and, therefore, demands the estimation of the RHFA for other gold isotopes far from stability in order to obtain their $g$ factors reliably. At the same time, it is the RHFA that became the source of the largest uncertainty in the hyperfine splitting calculations which are necessary to extract the effects of the parity and time-reversal symmetries violation of fundamental interactions in atom (see~\cite{skripn20,roberts21} and references therein).

The method of RHFA determination based on the analysis of $A$-constants ratio for different atomic states using atomic calculations was proposed in Refs.~\cite{ehlers68,persson98} and applied for several thallium~\cite{b2012} and bismuth~\cite{schmidt2018} isotopes.

When RHFA value is estimated and the ratios of the hyperfine constants were measured for different isotopes, we can obtain $g$ factors for the short lived isotopes ($g^*_I$) by the relation:
\begin{align} \label{g_unst}
g^*_I = g^\mathrm{ ref}_I \frac{A^*}{A^\mathrm{ref}}\left ( 1+ {}^\mathrm{ref}\Delta^*\right).
 \end{align} 
 Thus, the method of the RHFA evaluation enables one to increase the accuracy of the $g$ factor determination and, at the same time, to trace the isotopic change of the magnetization distribution inside the nucleus.

The aim of the present work is to describe in details the atomic calculations needed for the RHFA determination from experimental data on HFS for different atomic levels in gold. In comparison with our previous theoretical results briefly outlined in Ref.~\cite{B2020}, we significantly extended the basis sets and improved saturation of the correlation corrections.  We also calculated HFS constants for the Au-like ion of mercury and compared them with the experiment. We additionally made trial configuration-interaction (CI) calculation of gold as an eleven-electron system to estimate the possible influence of the $5d$-electrons excitations. As the result we managed to estimate the accuracy of our atomic calculations. This accuracy proved to be sufficient to draw tentative conclusions about the purity of the nuclear configurations in some gold isotopes with spin $I = 3/2$.
%%%%%%%%%%%%%%%%%%%%%%%%%%%%%%%%%%%%%%%%%%%%%%%%%%%%%%%%
\section{Basic formulas}
\subsection{Finite nuclear size corrections to HFS constants}
The $A$ constant for the finite nucleus can be written in the following form \cite{stroke61}:
\begin{align}\label{HFS_param}
    A &= g_I {\cal A}_0 (1-\delta)(1-\varepsilon)\,,
\end{align} 
where $g_I {\cal A}_0$ is the HFS constant in the case of a point-like nucleus, $\delta$ and $\varepsilon$ are the nuclear charge-distribution (Breit--Rosenthal, BR)~\cite{RB32,CS49} and magnetization-distribution (Bohr--Weisskopf, BW)~\cite{BW50} corrections, respectively.

In our atomic calculations nucleus is considered as a homogeneously charged and magnetized sphere of the radius \mbox{$R = \sqrt{5/3}\, r_\mathrm{rms}$}, where \mbox{$r_\mathrm{rms}=\langle r^2\rangle^{1/2}$} is a root-mean-square nuclear charge radius.

In terms of this model BR correction can be written as~\cite{ionesco60,Sha94}: 
\begin{align}\label{BR-bN}
\delta &= \delta(R) = b_N (R/\lambdabar_C)^\varkappa\,, 
    \quad \varkappa = 2\sqrt{1-(\alpha Z)^2}-1\,.
\end{align}
Here $\lambdabar_C$ is the Compton wavelength of the electron $\lambdabar_C =\hbar/(m_e c)$, $\alpha$ is the fine structure constant, $Z$ is nuclear charge, and dimensionless parameter $b_N$ depends on the electronic wavefunction of the atom and has to be calculated numerically. 

The BW correction $\varepsilon$ depends on the distribution of magnetization inside the nucleus and can be markedly different for various isotopes.
It was shown in Refs.~\cite{BW50,schwartz55} that the BW correction may be represented in the single-particle nuclear shell model 
and one-electron approximation as the product of two factors, one of them being dependent only on the atomic structure, the second being dependent only on the nuclear properties. This atomic-nuclear factorization was confirmed in Refs.~\cite{fujita75,MP95}, where more-refined atomic and nuclear models were used. Correspondingly, we introduced the nuclear factor $d_\mathrm{nuc}$ for parameterization of nuclear effects~\cite{KKDB17}:  
\begin{align} \label{BW-bM}
    \varepsilon = \varepsilon(R, d_\mathrm{nuc}) = b_M d_\mathrm{nuc} (R/\lambdabar_C)^\varkappa,
\end{align}
The nuclear factor is defined so that $d_\mathrm{nuc}=0$ corresponds to the point-like magnetic dipole in the center of the nucleus and $d_\mathrm{nuc}=1$ corresponds to the homogeneously magnetized sphere of radius $R$. The parameter $b_M$ has to be calculated numerically.

In terms of BR and BW corrections the RHFA for isotopes (1) and (2) defined by~\Eref{HFAdef} may be represented as:
\begin{align}\label{HFAdef1}
        ^1\Delta^2 \approx  \delta^{(2)} + \varepsilon^{(2)}-\delta^{(1)}- \varepsilon^{(1)}.
\end{align}
%%%%%%%%%%%%%%%%%%%%%%%%%%%
\subsection{Differential hyperfine anomaly}
%%%%%%%%%%%%%%%%%%%%%%%%%%%
For a number of isotopic sequences the ratios of the HFS constants for low-lying $s_{1/2}$ and $p_{1/2}$ atomic states \mbox{$\rho = A(p_{1/2})/A(s_{1/2})$}
were measured with sufficient accuracy to determine reliably the differential hyperfine anomaly (DHFA), defined as:
%$^{1}_{s_{1/2}}\Delta^{2}_{p_{1/2}}$:
\begin{align}\label{dhfa}
	\prescript{1}{p_{1/2}}\Delta^{2}_{s_{1/2}} = \frac{\rho^{(1)}}{\rho^{(2)}}-1 & = \frac{1+{}^1\Delta^2_{p_{1/2}}}{1+{}^1\Delta^2_{s_{1/2}}}-1\approx \nonumber \\ 
	& \approx {}^1\Delta^2_{p_{1/2}} - {}^1\Delta^2_{s_{1/2}}.
\end{align}
In the case of gold isotopes the BR correction is much smaller than the BW one (see below). 
Neglecting the BR correction, one can calculate the ratio of the relative hyperfine anomalies [see \Eref{BW-bM}]  by using the relation:
\begin{align}\label{ratio}
        \frac{{}^1\Delta^2_{s_{1/2}}}{{}^1\Delta^2_{p_{1/2}}} \approx \frac{b_{M\,s_{1/2}}}{b_{M\,p_{1/2}}}\equiv\eta.
\end{align}
Thus, the $\eta$ factor is determined solely by the electronic wave function. 
%properties of the atom.

With calculated atomic factor $\eta$ one can use experimental value of DHFA to determine relative magnetic hyperfine anomaly:
\begin{align}\label{mu_corr}
        {}^{1}\Delta^{2}_{s_{1/2}} &= \frac{\prescript{1}{p_{1/2}}\Delta^{2}_{s_{1/2}}}{1/\eta-1-\prescript{1}{p_{1/2}}\Delta^{2}_{s_{1/2}}}
\end{align}
and find nuclear $g$ factor using \Eref{g_unst}. 

Equations \eqref{HFAdef} and \eqref{BW-bM} enable one to determine $d_\mathrm{nuc}^{(2)}$ for isotope (2) provided the nuclear factor $d_\mathrm{nuc}^{(1)}$ for reference isotope and RHFA value ${}^{1}\Delta^{2}_{s_{1/2}}$ are known:
\begin{align}\label{dhfa2}
        d_\mathrm{nuc}^{(2)} = \frac{{}^{1}\Delta^{2}_{s_{1/2}}+d_\mathrm{nuc}^{(1)}b_{M\,s_{1/2}} (R/\lambdabar_C)^\varkappa} {\left(1+{}^{1}\Delta^{2}_{s_{1/2}}\right)b_{M\,s_{1/2}} (R/\lambdabar_C)^\varkappa}.
\end{align}

%%%%%%%%%%%%%%%%%%%%%%%%%%%%%%%%%%%%%%%%%%%%%%%%%%%%%%%%
\subsection{Parameterization of HFS constants}
In order to perform atomic calculation of the $A$ constants, we should specify three nuclear parameters:
\begin{align}\label{HFS_nuc}
    A &= A(g_I,d_\mathrm{nuc},R)\, =\,g_IA(1,d_\mathrm{nuc},R)\, .
\end{align}
To find parameter $b_M$ we calculated $A$-constants for the point-like magnetic dipole ($d_\mathrm{nuc}=0$) and for the homogeneously magnetized sphere ($d_\mathrm{nuc}=1$). Then $b_M$ is given by:  
\begin{align}\label{b_M}
	b_M &= (R/\lambdabar_C)^{-\varkappa} 
    \left(1 - 
    \frac{A(1,1,R)}{A(1,0,R)}
    \right)\,.
\end{align}
To find parameter $b_N$ we made calculations for different nuclear radii $R$: 
\begin{align}\label{b_N}
    b_N &= 
	\frac{\left(A(1,0,R_2) 
	- A(1,0,R_1)\right)
    \lambdabar_C^{\varkappa}}
    {A(1,0,R_2)R_1^{\varkappa}
    -A(1,0,R_1)R_2^{\varkappa}}
    \,.
\end{align}
After that the atomic parameter ${\cal A}_0$ was found from the relation:
\begin{align}\label{A_0}
    {\cal A}_0 &= 
    \frac{A(1,0,R)}
	{1 - b_N (R/\lambdabar_C)^{\varkappa}} 
    \,.
\end{align}
\subsection{Analytical estimates for the ratio of relative hyperfine anomalies}
Combining Eqs. \eqref{ratio} and \eqref{b_M}, one obtains:
\begin{align}\label{eta_frac}
	\frac{1}{\eta} = & \frac{A_{s_{1/2}}(1,0,R)}{A_{p_{1/2}}(1,0,R)} \frac{A_{p_{1/2}}(1,0,R)-A_{p_{1/2}}(1,1,R)}{A_{s_{1/2}}(1,0,R)-A_{s_{1/2}}(1,1,R)}.
\end{align}
If the principal quantum numbers for both electronic states are the same, then neglecting a small BR correction the first fraction in \Eref{eta_frac}  is ${\cal A}_{0,\,s_{1/2}}/{\cal A}_{0,\,p_{1/2}}\approx 3$~\cite{SFK78}. The second fraction depends on the radial integrals inside the nucleus of radius $R$, where the wave functions of $s_{1/2}$ and $p_{1/2}$ states are proportional to each other with proportionality coefficient $\frac{\alpha^2 Z^2}{4}\left (1+\frac{\alpha^2 Z^2}{4} \right)^2$.
Thus, the leading-order term for $1/\eta$ is independent of the principal quantum number \cite{ita20}:
\begin{align}\label{eta_first_order}
        \frac{1}{\eta} = \frac{3}{4} \alpha^2 Z^2.
\end{align}
For the atoms considered here it gives $\eta (\mathrm{Au}) = 4.0$ and $\eta (\mathrm{Hg}) = 3.9$.

\section{Calculation results}
We consider the members of the gold isoelectronic sequence (Au I and Hg II) as one-electron systems, with the $5d^{10}$ electron shell included in the atomic core. Correlation between the single and all core electrons is treated perturbatively. In this case only the lower part of the valence spectrum can be represented with reasonable accuracy. 
For energies higher than the core excitation energy, the effective Hamiltonian has poles and the results may be unreliable. In the case of gold the $5d^{10}6p$ doublet lies on the borderline of the core-excitations. Interaction between this doublet and the quadruplet $5d^9_{5/2}6s_{1/2}6p_{3/2}\, {}^4P^o$ is rather strong. Due to the opposite order of levels in these multiplets, the energy gap between levels with $J=3/2$ is much smaller, than the gap between $J=1/2$ levels. Therefore, the inaccuracy in the description of the $5d^{10}6p_{3/2}$ state in gold is expected to be markedly larger, than that of the $5d^{10}6p_{1/2}$ level. Correspondingly, below we restrict ourselves only to $6s_{1/2}$ and $6p_{1/2}$ states.

All calculations are performed using either Dirac-Coulomb, or Dirac-Coulomb-Breit Hamiltonians. 
%Breit corrections are taken into account beyond the Gaunt term~\cite{breit1,breit2}. 
Breit corrections including both the magnetic term and the retardation term in the zero-frequency limit are taken into account in accordance with Refs.~\cite{breit1,breit2}. When the Dirac-Coulomb-Breit Hamiltonian is used we mark the corresponding results by addition the notation ``+Breit'' to the designation of the calculation method. We start by solving Dirac-Hartree-Fock (DHF) equations for the core and valence orbitals up to $7p_{3/2}$. 
After that we merge these orbitals with B-splines of order 4 as described in Ref.~\cite{basis} to form a basis set for calculating the correlation corrections. 
The basis set $22spdfghi$ includes 224 orbitals for partial waves with orbital angular momentum $l$ from 0 to 6.

Correlation corrections to the HFS include corrections to the many-electron wave functions and to the hyperfine operator. We use either second order many-body perturbation theory (MBPT) \cite{DFK96b,KPST15}, or linearized single double couple clusters method (LCC) \cite{Koz04, SKJJ09} to take into account correlation corrections to the wave function. Both MBPT and LCC corrections include self-energy contribution \cite{DFKP98}. The energy dependence of the effective Hamiltonian is taken into account as discussed in Refs.\ \cite{DFK96b,SKJJ09}. To account for the correlation corrections to the hyperfine operator we use the random phase approximation (RPA) with structural radiation correction \cite{KPJ01}. These corrections include in particular the spin polarization of the core shells, up to $1s$. 

\begin{table}[tbh]
\caption{\label{tbl:energies}
Binding energies and their difference (in au) for the low-lying states of the Au I and Hg II within one electron approach.
The rows DHF, MBPT, and LCC correspond to the Dirac--Hartree--Fock, Dirac--Hartree--Fock plus MBPT, and Dirac--Hartree--Fock plus linearized coupled clusters methods, respectively. We also give Breit corrections to the energy in the LCC approximation.
The final theoretical values (LCC+Breit), experimental data, and the theoretical error (in \%) are listed in the last three rows for each system.}
\begin{tabular}{lcccc}
\hline
\\[-3mm]
Method                  &$6s_{1/2}$&$6p_{1/2}$& Diff.\\
\hline
\multicolumn{4}{c}{ Au I}\\
DHF                  &{~~0.2746~~}&{~~0.1338~~}&{~~0.1408~~}\\
MBPT                 &{~~0.3513~~}&{~~0.1703~~}&{~~0.1811~~}\\
LCC                  &{~~0.3402~~}&{~~0.1694~~}&{~~0.1708~~}\\
LCC+Breit               &{~~0.3397~~}&{~~0.1689~~}&{~~0.1708~~}\\
Expt~\cite{NIST}     &{~~0.3390~~}&{~~0.1688~~}&{~~0.1702~~}\\
Diff.with expt.      &{$-0.19$\%} &{~$-0.05$\%}&{~$-0.32$\%}\\
\multicolumn{4}{c}{Hg II} \\
DHF                  &{~~0.6218~~}&{~~0.4087~~}&{~~0.2131~~}\\
MBPT                 &{~~0.7041~~}&{~~0.4610~~}&{~~0.2431~~}\\
LCC                  &{~~0.6912~~}&{~~0.4561~~}&{~~0.2351~~}\\
LCC+Breit               &{~~0.6905~~}&{~~0.4553~~}&{~~0.2352~~}\\
Expt~\cite{NIST}     &{~~0.6893~~}&{~~0.4547~~}&{~~0.2346~~}\\
Diff.with expt.      &{$-0.17$\%} &{$-0.12$\%}&{$-0.26$\%} \\
\hline 
\end{tabular}
\end{table}

\begin{table*}[tbh]
\caption{\label{tbl:hfs}
The atomic parameters ${\cal A}_0$ (MHz), $b_N$, $b_M$, and HFS constants for the lower levels of gold and Au-like mercury ion.
We compare HFS constants for $^{193}\mathrm{Au}^m$ (with $d_\mathrm{nuc} = 0.73$) and $\rm ^{199}Hg$ (with $\varepsilon_{6s_{1/2}} = 2.0\%$ and $\varepsilon_{6p_{1/2}} = \frac{\varepsilon_{6s_{1/2}}}{\eta}$) with available experimental data~\cite{B2020,hgII_s,hgII}.
}
\begin{tabular}{lccccccccc}
\hline
\\[-3mm]
        \multirow{3}{*} {Method} &\multicolumn{4}{c}{$6s_{1/2}$} &\multicolumn{4}{c}{$6p_{1/2}$}&\\
         \cmidrule(lr){2-5} \cmidrule(lr){6-9} 
         &\multicolumn{1}{c}{ ${\cal A}_0$ (MHz)}&\multicolumn{1}{c}{$b_N$}&\multicolumn{1}{c}{$b_M$}&\multicolumn{1}{c}{$A$  (MHz)}
         &\multicolumn{1}{c}{${\cal A}_0$ (MHz)}&\multicolumn{1}{c}{$b_N$}&\multicolumn{1}{c}{$b_M$}&\multicolumn{1}{c}{$A$  (MHz)}
         &\multicolumn{1}{c}{$\eta$} \\
\hline
\multicolumn{10}{c}{Au I}\\
DHF                  &24435&{1.260~~}&0.244   & 24944 &2166  &{0.337~~}&0.065   & 2414&{~~3.74}\\
DHF+Breit            &24362&{1.259~~}&0.244   & 24872 &2140  &{0.336~~}&0.065   & 2385&{~~3.74}\\
RPA                  &26851&{1.260~~}&0.242   & 27401 &1848  &{0.257~~}&0.058   & 2074&{~~4.21}\\
RPA+Breit            &26797&{1.263~~}&0.242   & 27351 &1833  &{0.257~~}&0.057   & 2057&{~~4.23}\\
RPA+MBPT             &34349&{1.260~~}&0.240   & 35073 &3424  &{0.292~~}&0.060   & 3831&{~~3.99}\\
RPA+MBPT+Breit       &34285&{1.258~~}&0.239   & 35014 &3395  &{0.291~~}&0.060   & 3799&{~~3.99}\\
RPA+LCC              &32472&{1.259~~}&0.240   & 33158 &3207  &{0.296~~}&0.060   & 3587&{~~3.99}\\
RPA+LCC+Breit        &32411&{1.257~~}&0.239   & 33101 &3180  &{0.296~~}&0.060   & 3558&{~~4.00}\\
Experiment ($^{193}\mathrm{Au}^m$)&&&&\multicolumn{1}{r}{~~~~~32391(30)}&&&&\multicolumn{1}{r}{~~~3696(4)}&\\
Relative error       &&&&2.2\%&{~~}&&&$-3.7$\%  &{~~}\\
\hline
\multicolumn{10}{c}{Hg II}\\
DHF                  &38286&{1.298~~}&0.253&34063&5661  &{0.357~~}&0.070&5535&{~~3.62}\\
DHF+Breit            &38197&{1.297~~}&0.253&33987&5605  &{0.356~~}&0.070&5481&{~~3.63}\\
RPA                  &41474&{1.297~~}&0.251&36910&5725  &{0.330~~}&0.066&5612&{~~3.79}\\
RPA+Breit            &41414&{1.295~~}&0.251&36864&5681  &{0.329~~}&0.066&5569&{~~3.80}\\
RPA+MBPT             &47012&{1.267~~}&0.248&41960&7402  &{0.340~~}&0.067&7249&{~~3.71}\\
RPA+MBPT+Breit       &46942&{1.265~~}&0.247&41907&7347  &{0.339~~}&0.067&7196&{~~3.72}\\
RPA+LCC              &45410&{1.270~~}&0.248&40518&7050  &{0.341~~}&0.067&6903&{~~3.71}\\
RPA+LCC+Breit        &45342&{1.268~~}&0.248&40474&6998  &{0.340~~}&0.067&6853&{~~3.72}\\
%Experiment ($\rm ^{199}Hg$)&&&&\multicolumn{3}{l}{~~40507.3479968416(4)}&&\multicolumn{2}{l}{~~~6970(90)}\\
Experiment ($\rm ^{199}Hg$)&&&&\multicolumn{3}{l}{~~~~~40507.3479968416(4)}&&\multicolumn{2}{l}{~~~6970(90)}\\
Relative error       &&&&$-0.08$\%&{~~}&&&$-1.7$\%  &{~~}\\
\hline 
\end{tabular}
\end{table*}
 In \Tref{tbl:energies} the calculated binding energies for the low-lying states of Au I and Hg II within one electron approach are compared with the experimental data~\cite{NIST}. In \Tref{tbl:hfs} our results for $A$-constants of gold and Au-like mercury ion are summarized.
As is clearly seen from Tables I and II, the LCC method in all cases gives results which are closer to the experimental binding energies than for the MBPT method.

The factor $\eta (\mathrm{Au})$ equals 3.74 in the DHF approximation. In the next step we take into account effective mixing of different partial waves via RPA corrections. The factor $\eta (\mathrm{Au})$ increases due to these corrections up to 4.23. Electron correlation corrections taken into account within MBPT, or LCC methods significantly change ${\cal A}_0$ values. The parameters $b_N$ and $b_M$ are also changed due to the structural radiation correction, since this correction, like RPA, mixes different partial waves. Contributions of the RPA and the structural radiation corrections to the $b_N$ and $b_M$ parameters tend to nearly cancel each other. Both these contributions  to  the  $b_N$ and $b_M$ parameters  appear to be rather large, which leads  to  the  instability  of  the  results. At the same time, the results for these parameters are almost the same for the MBPT and the LCC approximations.

Calculation of the parameters $b_N$ require a small change in the nuclear radius, which leads to a change of integration grid. Therefore, the parameters $b_N$ are more sensitive to the size of the basis set. In comparison with our previous theoretical results presented in Ref.~\cite{B2020} we, in particular, significantly extended the basis set in order to check the stability of the results. Because of that our  present  values  for  the $b_N$ parameters  differ from  those  in  Ref.~\cite{B2020}  by  20\%,  while  the  $b_M$ values  remained  practically  unchanged.

As a final value for the $\eta$ factor in gold we adopt the mean value of the results obtained in the frameworks of the different approximations (see the last column in \Tref{tbl:hfs}) with the uncertainty covering the dispersion of these results: $\eta (\mathrm{Au})=4.0(3)$.
The new value coincides with the value calculated in Ref.~\cite{B2020}. Note, that the new $\eta (\mathrm{Au})$ value differs noticeably from the corresponding values obtained and used previously: 3.2~\cite{BW50}, 3.5~\cite{eisinger}, 4.5~\cite{stroke61}, 3.3~\cite{SWYG07}.

Within the method discussed above the interaction of the valence electron with ten $5d$ electrons was treated perturbatively. We check the reliability of this approach by the CI calculation for the eleven outermost electrons. Within this approach all interactions with the $5d$ electrons are taken into account non-perturbatively. Because of the computational limitations we have to use a much shorter basis set $10sp9d8fgh$ for the eleven-electron CI. 
The $b_M$ parameters for the $6s_{1/2}$ state of gold obtained within CI (11e)+Breit and RPA+LCC+Breit approaches differ from each other by 2\%, while for the $6p_{1/2}$ state the similar difference is 5\%. One important effect which is missing in the CI calculation is the mixing between $s$ and $p$ partial waves since all inner $s$ and $p$ shells are kept frozen. This mixing primarily affects smaller HFS constants, which explains better agreement for the constant $A(6s_{1/2})$. As the result, $\eta$ (CI[11e]+Breit) = 4.28 differs from our final value by ~6\% and is covered by the ascribed uncertainty. This means that in our RPA+LCC+Breit calculation we did not miss any significant higher-order contributions from the correlations with the $5d$ shell at least for the $\eta$-factor calculation.

We also calculated the $\eta$ factors for Hg II within the one-electron approach. Contributions of the RPA and structural radiation corrections are smaller for this ion. At the same time, excitation energies from the $5d$ electron shell are larger. Correspondingly, dispersion of the different-approximation results for $\eta$ value is significantly less for Hg II than that for Au I and uncertainty for $\eta$(Hg II) is smaller. Final value for $\eta (\mathrm{Hg\, II})$ is equal to 3.71(9).
%%%%%%%%%%%%%%%%%%%%%%%%%%%%%%%%%%%%%%%%%%
\section{The nuclear factor}
%%%%%%%%%%%%%%%%%%%%%%%%%%%%%%%%%%%%%%%%%%
In the framework of the single-particle nuclear model for odd-$A$ nuclei~\cite{BW50,B51} the BW correction $\varepsilon$ is the linear combination of the spin, $\varepsilon_S$, and orbital, $\varepsilon_L$, contributions:
 \begin{align}
 \label{BW_1}
	 \varepsilon = \alpha_S \left [\varepsilon_S +\zeta\left (\varepsilon_S-\varepsilon_L \right)\right ] + (1-\alpha_S)\varepsilon_L.
 \end{align}
 Coefficient $\alpha_S$ is the fraction of the spin contribution to the nuclear magnetic moment~\cite{BW50}: 
\begin{align}
 \label{alphas}
        &\alpha_S = \frac{g_S}{g_I} \frac{g_I - g_L}{g_S- g_L},
\end{align}
where $g_S$ and $g_L$ are spin and orbital $g$-factors respectively.
The nuclear $g$-factor is given by the famous Land{\'e} formula:
\begin{align}
\label{g_S}
g_I = \frac{1}{2}\left [ (g_L+g_S) + (g_L - g_S)\frac{L(L+1)-3/4}{I(I+1)} \right], 
\end{align}
where $L$ is the orbital momentum  of the valence shell-model state. In case of a valence proton $g_L = 1$, while for neutron $g_L = 0$; $g_S$ is chosen  from the condition that \Eref{g_S} reproduces experimental $g$-factor value. The parameter $\zeta$ in \Eref{BW_1} takes into account the angular asymmetry of the spin distribution and is given by~\cite{B51}:
\begin{align}
 \label{zeta}
\zeta =
 \begin{cases}
        \frac{2I-1}{4(I+1)} & \text{if $I = L+\frac{1}{2}$} \\ 
         \frac{2I+3}{4I} & \text{if $I = L-\frac{1}{2}$}. \\
 \end{cases}
\end{align}

Within the model of  homogeneously magnetized sphere of radius $R$ the spin contribution $\varepsilon_S$ is given by the relation~\cite{KKDB17}:
 \begin{align}
 \label{BW}
	 \varepsilon_S = b_M (R/\lambdabar_C)^\varkappa.
 \end{align}
Comparing Eqs. \eqref{BW-bM}, \eqref{BW_1} and \eqref{BW} one obtains the expression for $d_\mathrm{nuc}$ in the single-particle model for the odd-$A$ nuclei:
 \begin{align}
 \label{d_nuc}
         d_\mathrm{nuc} = ( 1 + \zeta) \alpha_S + \frac{\varepsilon_L}{\varepsilon_S} \left[1 - \alpha_S ( 1 + \zeta)\right].
 \end{align}
According to Ref.~\cite{BW50} the value of $\varepsilon_L/\varepsilon_S$ is equal to 0.62.

Note the variety of models that are currently used to describe the nuclear magnetization distribution, e.g. the surface-current nuclear model proposed in Ref.~\cite{Sliv51} and developed in detail in Ref.~\cite{Kar15}.

We consider $11/2^-$ and $3/2^+$ long-lived states in odd-$A$ gold isotopes as single-hole states $(\pi h_{11/2})^{-1}$ and $(\pi d_{3/2})^{-1}$ respectively ($\pi$ denotes the valence proton). Nuclear factors calculated by Eqs.~(\ref{alphas} -- \ref{d_nuc}) with these assumptions, are presented in \Tref{tbl:g_factors}.
This approximation is evidently simplistic since the corresponding nuclei are known as weakly deformed~\cite{sauvage00}. 
However, it will be shown that even such a rough approximation enables one to explain the trends in the observed RHFA. 

%%%%%%%%%%%%%%%%%%%%
\begin{table}[th]
\caption{
Comparison of the factors $d_\mathrm{nuc}$ obtained by Eqs.~(\ref{alphas} -- \ref{d_nuc}) and \Eref{dhfa2} with $^{193}\mathrm{Au}^m$ as the reference isotope, respectively. Spins and magnetic moments of the corresponding isotopes or isomers are also shown.\label{tbl:g_factors}}
\begin{tabular}{lclccc}
\hline
	Isotope& Nuclear &\multicolumn{2}{c}{$g$-factor} &\multicolumn{2}{c}{$d_\mathrm{nuc}$}\\
           & spin    &         &Ref.&Eqs.~(\ref{alphas} -- \ref{d_nuc})&\Eref{dhfa2}\\
\hline
	199& 3/2& 0.1799(5)&\cite{au199}&$-3.7$&$-3.2(5)$\\
	197& 3/2& 0.097164(6)&\cite{dahmen67} &$-8.0$&$-5.1(5)$\\
	195& 3/2& 0.0991(4)&\cite{au195}&$-7.8$&--\\
	193& 3/2& 0.0932(10)&\cite{B2020}&$-8.4$&$-5.4(8)$\\
	191& 3/2& 0.0908(12)&\cite{B2020}&$-8.6$&$-5.8(9)$\\%0.0913(6)
	195$^m$& 11/2& 1.148(7)&\cite{B2020}&0.73&\\%\multirow{5}{*}{$\Vast \} 0.73$}\\
	193$^m$& 11/2& 1.149(7)&\cite{B2020}&0.73&\\
	191$^m$& 11/2& 1.150(7)&\cite{B2020}&0.73&\\
	189$^m$& 11/2& 1.157(7)&\cite{B2020}&0.73&\\
	177$^m$& 11/2& 1.185(7)&\cite{B2020}&0.74&\\
\hline
\end{tabular}
\end{table}
%%%%%%%%%%%%%%%%%%%%%%%%%%%%%%%%%%%%%%%%%%%%%%%%%%%%%%%%%%%%%%%
\section{Accuracy of the atomic calculations}
%%%%%%%%%%%%%%%%%%%%%%%%%%%%%%%%%%%%%%%%%%%%%%%%%%%%%%%%%%%%%%%
In order to estimate the reliability of the nuclear factor evaluation for gold isotopes by \Eref{d_nuc}, one needs to assess the accuracy of the calculation of all other parameters. To this end, let us consider mercury ions (Hg II) with the atomic structure which is very similar to that of the gold atom and where BW corrections were studied previously in more detail. Experimental values of HFS constants were determined in Refs.~\cite{hgII_s, hgII_201}: $A$($6s_{1/2}$, $\rm ^{199}Hg\,\, II) = 40507.348$ MHz and $A$($6s_{1/2}$, $\rm ^{201}Hg\,\, II) = -14977.183$ MHz.
One can compare with experiment only the $A$ values, whereas our main goal is to estimate the accuracy of the ${\cal A}_0$ calculations. 
Uncertainty of the theoretical HFS constants consists of three parts: uncertainties of ${\cal A}_0$, $\delta$, and $\varepsilon$ [see \Eref{HFS_param}]. Thus, one should estimate independently the accuracy of $\delta$ and $\varepsilon$ determination in order to draw conclusions about the accuracy of the ${\cal A}_0$ calculations from the comparison with experiment. The accuracy of the BR correction ($\delta$) calculation can be estimated as 3\% from the dispersion of the results of different approximations in $b_N$ calculation (see \Tref{tbl:hfs}). This leads to the uncertainty of 0.3\% in the final $A(6s_{1/2})$ values. 

To estimate the accuracy of BW correction ($\varepsilon$) a semiempirical approach was used. It was shown by \citet{ML73} that the relative hyperfine anomalies for a series of mercury isotopes, including $\rm ^{201,\, 199}Hg$, are well reproduced assuming the following simple relation (ML rule):
\begin{align}
 \label{ML_rule}
         \varepsilon = \pm \frac{\alpha}{\mu}, \quad I = L \pm \frac{1}{2}, \quad \alpha = 1 \cdot 10^{-2}.
 \end{align}
 where $\mu$ is the magnetic moment of the isotope in question. 
 %Note that the ML rule proves to be valid only for the mercury isotopes and it does not work for the gold nuclei~\cite{B2020}. 
 Using the known $\mu$ values~\cite{stone05}, one obtains: $\varepsilon$($\rm ^{199}Hg$) = 2.0\%, $\varepsilon$($\rm ^{201}Hg$) = 1.8\%. Correspondingly, theoretical $A(6s_{1/2})$ constants for $\rm ^{201,\, 199}Hg$ calculated with these BW corrections and ${\cal A}_0$, $b_N$ from \Tref{tbl:hfs} are $-14978$~MHz and $40474$~MHz, respectively, which means relative error (in comparison with the experimental data) 0.06\% and 0.08\%.
However, it does not mean with certainty that the accuracy of the ${\cal A}_0$ calculation is of the same order.
We estimated the possible variations of the $\varepsilon$ correction which might lead to the uncertainty in the ${\cal A}_0$ value using the theoretical justification of the ML rule by \citet{fujita75}. They gave the following expression for BW correction in mercury:
\begin{align}
 \label{ML_rule2}
         \varepsilon = c_1 \pm \frac{\alpha}{\mu},
 \end{align}
where constant $c_1$ is less than 0.02 for mercury nuclei (see also~\cite{ekstrom80}). The RHFA description is the same for both $\varepsilon$ representations [Eqs.~(\ref{ML_rule}) or (\ref{ML_rule2})], however, absolute $\varepsilon$ values may be different. This possible difference was accepted as the measure of the BW corrections uncertainty. This uncertainty results in the 2\% error in the final $A(6s_{1/2})$ values. Consequently, taking into account the excellent agreement of the theoretical and experimental $A(6s_{1/2})$ values for $\rm ^{201,\, 199}Hg^+$, one can conservatively estimate the possible uncertainty of the ${\cal A}_0 (6s_{1/2})$ calculation for Hg II as ~ 2.5\%. Keeping in mind the similarity of the atomic structure in Au I and Hg II, one can expect that the same estimation is valid for gold atom.
%%%%%%%%%%%%%%%%%%%%%%%%%%%%%%%%%%%%%%%%
\section{Calculation of nuclear factors}
%%%%%%%%%%%%%%%%%%%%%%%%%%%%%%%%%%%%%%%%
For $6s_{1/2}$ state in $^{193}\mathrm{Au}^m$, we deduced $d_\mathrm{nuc} = 0.73$ by \Eref{d_nuc}. This factor corresponds to the BW correction of $\varepsilon=1.4\%$ [see \Eref{BW-bM}]. Experimentally measured $A (6s_{1/2},\,\, ^{193}\mathrm{Au}^m )$ agree with our final result within 2.2\%. At the same time, $A (6s_{1/2},\,\, ^{197}\mathrm{Au}) = 3265$~MHz calculated with the single-particle value [\Eref{d_nuc}] $d_\mathrm{nuc}=-8.0$, differs from the experimental value  3049.66~MHz~\cite{Au_stable_s} by 7.1\%. This difference can not be attributed to the inaccuracy of the atomic calculations, since we estimated this inaccuracy to be less than 2.5\%. This means that single-particle model does not work for $d_\mathrm{nuc}$ calculation in $\rm ^{197}Au$. At the same time, the agreement between theoretical and experimental values for $A(6s_{1/2},\, ^{193}\mathrm{Au}^m)$ indicates that the single-particle $d_\mathrm{nuc}$ value for $^{193}\mathrm{Au}^m$ does not contradict the available experimental data and uncertainty estimation for the atomic calculations. Therefore, we tried to extract information about the nuclear factor in $\rm ^{197}Au$ with the aid of \Eref{dhfa2} and $d_\mathrm{nuc} (^{193}\mathrm{Au}^m)$. It should be noted that the nuclear factor $ d_\mathrm{nuc}$ has singularity when $g_I\to 0$ [see~\Eref{alphas}]. 
%%%%%%%%%%%%%%%%%%%%%%%%%%%%%%%%%%%%%%%%%%%%%%%%%%%%%%%%%%%%%%%
%\section{Extraction of the nuclear factors from experimental data}
%%%%%%%%%%%%%%%%%%%%%%%%%%%%%%%%%%%%%%%%%%%%%%%%%%%%%%%%%%%%%%%
We used $d_\mathrm{nuc}(^{193}\mathrm{Au}^m)= 0.73$ obtained within the single-particle nuclear model to restore $d_\mathrm{nuc}^{197}$ from ${}^{197}\Delta^{193^m}(6s_{1/2}) = 11.2(11)\%$~\cite{B2020} by \Eref{dhfa2}: $d_\mathrm{nuc}^{197}=-5.1(5)$. 
The uncertainties in $d_\mathrm{ nuc}$, calculated by \Eref{dhfa2}, are determined by the uncertainty of $\eta$ factor and experimental uncertainties of the corresponding RHFS values and do not include the possible error of the reference $d_\mathrm{nuc}(^{193}\mathrm{Au}^m)$ value. Equation \eqref{BW-bM} gives the BW correction $\varepsilon^{197}=-9.6(9)\%$ for $6s_{1/2}$ state. We obtained \mbox{$d_\mathrm{ nuc}^{193} = -5.4(8)$} and $d_\mathrm{ nuc}^{191} = -5.8(9)$ from RHFA values \mbox{$|{}^{197}\Delta^{193}(s_{1/2})| \le1.5\%$} and \mbox{$|{}^{197}\Delta^{191}(s_{1/2})|\le 1.4\%$} given in Ref.~\cite{B2020}. The same procedure gives for $\rm ^{199}Au$ \mbox{$d_\mathrm{ nuc}^{199} = -3.2(5)$}. The BW correction for $6s_{1/2}$ state of $\rm ^{199}Au$ isotope is $\varepsilon^{199} = 6.0(9)\%$. This nuclear factor extracted from experimental data is in reasonable agreement with the prediction of the single-particle nuclear model [\Eref{d_nuc}]: \mbox{$d_\mathrm{nuc}^{199} = -3.7$}. At the same time, the factors for other gold 3/2 isotopes obtained within single-particle nuclear model are markedly overestimated (see~\Tref{tbl:g_factors}).
Thus one can assume that $3/2^+$ ground state of $\rm ^{199}Au$ belongs to the relatively pure $d_{3/2}$ configuration, whereas $3/2^+$ ground states of $\rm ^{191,\,193,\,195,\,197}Au$ have the significant admixture of other configurations, which leads to the discrepancy between single-particle and semiempirical [\Eref{d_nuc}] values of $d_\mathrm{nuc}$ factor.
 
We calculate HFS constants for $\rm ^{197}Au$ with known parameters ${\cal A}_0$, $b_N$, $b_M$, and \mbox{$g_I$ = 0.097164(6) $\mu_N$~\cite{dahmen67}} and $d_\mathrm{nuc}=-5.1$. Our final theoretical results $A(6s_{1/2},\, ^{197}\mathrm{Au}) = 3110$~MHz and $A(6p_{1/2},\, ^{197}\mathrm{Au}) = 309$~MHz agree with experimental values~\cite{Au_stable_s,Au_stable_p}  within 2.0\% and $-3.9$\%, respectively. 
%%%%%%%%%%%%%%%%%%%%%%%%%%%%%%%%%%%%%%%%%%
\section{Conclusions}
%%%%%%%%%%%%%%%%%%%%%%%%%%%%%%%%%%%%%%%%%%
The nuclear magnetic moments of the short-lived isotopes are extracted usually from the HFS constants. In this case the accuracy is limited by the relative hyperfine anomaly and it can be improved if we independently find the value of this anomaly. The latter can be done using the $A$-constant ratio for different atomic states and sufficiently accurate atomic calculations. To this end we here present more thorough calculations of the magnetic hyperfine structure constants of the  two lowest levels in neutral gold atom compared to the ones reported in Ref.~\cite{B2020}. We significantly extended the basis sets and improved saturation of the correlation corrections. We additionally made trial CI calculation of gold as an eleven-electron system to estimate the possible influence of the $5d$-electrons excitations. We also considered HFS of the Au-like mercury ion. Using~\citet{fujita75} representation of BW correction which was shown to explain the anomaly in a number of mercury isotopes,  one can estimate the possible uncertainty of the ${\cal A}_0 (6s_{1/2})$ calculation for both Au~I and Hg~II as 2.5\%. 

Experimentally measured $A(6s_{1/2},^{191}\mathrm{Au}^m)$ agrees with calculation results for single-particle nuclear model value $d_\mathrm{nuc}(^{191}\mathrm{Au}^m) = 0.73$, within 2.2\%. This indicates the applicability of this model for $^{193}\mathrm{Au}^m$ as well as for the other $11/2^-$ gold isomers. We used this $d_\mathrm{nuc}$ value
as a reference to restore the nuclear factors for other isotopes. The nuclear factor for the $\rm ^{199}Au$ obtained from the experimental data using our method, $d_\mathrm{ nuc}^{199} = -3.2(5)$, is in a reasonable agreement with the prediction of the single-particle nuclear model, $d_\mathrm{nuc}^{199} = -3.7$. The nuclear factors for $\rm ^{197,\,193,\,191}Au$ ($I = 3/2$) isotopes are close to each other.
From experimental data we found \mbox{$d_\mathrm{ nuc}= -5.5(6)$} for these three isotopes, whereas
the single-particle nuclear model gives $d_\mathrm{ nuc} = -8.2(4)$. One can assume that these isotopes have significant admixture of other configurations, whereas $\rm ^{199}Au$ belongs to the relatively pure $d_{3/2}$ configuration. %Note, that equations of the single-particle nuclear model %%%%%%%%%%%%%%%%%%%%%%%%%%%%%%%%%%%%%%%%%%
\subsection*{Acknowledgments}
%%%%%%%%%%%%%%%%%%%%%%%%%%%%%%%%%%%%%%%%%%
This research was funded  in part by the Russian Science Foundation Grant \textnumero 19-12-00157. A.E.B. acknowledges support by RFBR according to the research project \textnumero 19-02-00005. Calculations  were done on the Complex for Simulation and Data Processing for Mega-science Facilities at NRC “Kurchatov Institute”, http://ckp.nrcki.ru/. 
\bibliographystyle{apsrev}

\end{document}